\documentclass[sigconf]{acmart}
\usepackage{amsmath}
\usepackage{graphicx}
\usepackage{url}
\usepackage[font={small}]{subfig}

\newenvironment {squishlist}
{\begin{list}{$\bullet$}
  { \setlength{\itemsep}{0pt}
     \setlength{\parsep}{3pt}
     \setlength{\topsep}{3pt}
     \setlength{\partopsep}{0pt}
     \setlength{\leftmargin}{1.5em}
     \setlength{\labelwidth}{1em}
     \setlength{\labelsep}{0.5em} } }
{\end{list}}

\begin{document}

\acmConference{WebSci'17}{}{June 25-28, 2017, Troy, NY, USA.}
\title{Characterizing Fan Behavior to Study Para Social Breakups}
\titlenote{This is a longer version of a 2-page poster publised at WebSci'17. Please cite the WebSci poster, not the arxiv version.}

\fancyhead{}
\settopmatter{printacmref=false}

\author{Kiran Garimella}
\affiliation{%
  \institution{Aalto University}
  \city{Helsinki} 
  \country{Finland} 
}
\email{kiran.garimella@aalto.fi}

\author{Jonathan Cohen}
      \affiliation{
	\institution{University of Haifa}
	\city{Haifa}
	\country{Israel}
}
      \email{jcohen@com.haifa.ac.il}

\author{Ingmar Weber}
\affiliation{%
	\institution{Qatar Computing Research Institute}
  \city{Doha} 
  \country{Qatar}
}
\email{ingmarweber@acm.org}

\begin{abstract}
Celebrity and fandom have been studied extensively in real life. However, with more and more celebrities using social media, the dynamics of interaction between celebrities and fans has changed. 
Using data from a set of 57,000 fans for the top followed celebrities on Twitter, we define a wide range of features based on their Twitter activity. Using factor analysis we find the most important factors that underlie fan behavior.
Using these factors, we conduct analysis on (i) understanding fan behavior by gender \& age, and (ii) para-social breakup behavior.
We find that (i) fandom is a social phenomenon, (ii) female fans are often more devoted and younger fans are more active \& social, and (iii) the most devoted fans are more likely to be involved in a para-social breakup. 
Our findings confirm existing research on para-social interactions. 
Given the scale of our study and dependence on non-reactive data, our paper opens new avenues for research in para-social interactions.

\end{abstract}

\maketitle


\section{Introduction}

Since the advent of electronic media, para-social relationships (PSR) or para-social interactions (PSI) have been a widespread feature of advanced societies~\cite{horton1956mass}. 
Horton and Wohl define PSR as a one-way relationship that is imagined as a mutual relationship. They use the phrase `intimacy at a distance' to describe this phenomenon. It was used to explain the power of persuasion that certain celebrity personae have.
Following early studies of PSR based on self-report surveys, the use of Computer Mediated Communications has made the study of such phenomena through non-reactive measures possible~\cite{garimella2014love}. 
Traditionally, the celebrity-fan relationship was very lop-sided, with one way interaction between the celebrity and the fan (typically through the television). With the advent of social media, 
PSR have become more reciprocal and `social', or at least intensified the illusion of sociability by creating seemingly more personal and frequent communication between a celebrity and their fans.

Given this new era of PSI on social media, many studies have looked at how celebrity behavior has been shaped on social media~\cite{marwick2011see,stever2013twitter} and to a certain extent, how fandom has changed~\cite{stavros2014understanding}.
Eyal et al. \cite{eyal2006good} introduce the idea of a para-social breakup, the end of a PSR. They find that a para-social breakup is like a regular romantic relationship breakup but not as intense.
This para-social breakup (PSB) is manifested on social media by an act of unfollowing the celebrity.
Though breakups on social media have been studied in the past~\cite{garimella2014love}, para-social breakups have special characteristics and understanding them at a large scale would be of great interest, with applications in brand loyalty, advertising and societal good, in general. 

Most studies on PSR and PSB to date focus on self-report surveys, where fans report on their feelings, attitudes \& behaviors and are done on a small scale. We tackle such problems to build on top of these studies in this paper, and present a first large scale analysis of fan behavior on Twitter and how different fans indulge in PSR and PSB on Twitter.
The research questions we tackle in this study are:
(i) Can a systematic pattern be found in fan behaviors based on their interactions on Twitter?
(ii) Can such a typology reliably predict and help in understanding para social breakups with celebrities?

We collect data of 15 celebrities from popular culture and 57,000 of their fans on Twitter. We construct an exhaustive list of features from fan activity and fan-celebrity interaction on Twitter.
We then use factor analysis to automatically group these features into three main factors, that explain most of the fan interactions on twitter, namely, \emph{devotedness}, \emph{sociability}, and \emph{activity}. Using these factors as a building ground, we measure the effects of these factors on para-social breakup behavior of the fans
(identified using fans unfollowing celebrities on Twitter).

We make some interesting and novel observations. 
(i) Using inferred gender and age groups of the fans, we find that women are more devoted and younger fans tend to be more social \& active.
(ii) We find a correlation between how devoted a fan is and how social they are, 
which suggests that PSR on social media is a way to connect not only with the celebrity but with other fans and that this is an important component of fandom. Fans find other like-minded people of Twitter to connect with. This is in line with findings from~\cite{stavros2014understanding}.
(iii) We find that the more involved a fan is in a PSR, the more likely that they will end the PSR.
We propose a potential explanation for this finding using a cost/benefit analysis. 

To the best of our knowledge, this is the first study to characterise fan behavior on Twitter and  para-social breakups at such a large scale.
Our findings are valuable because of our dependence on non-reactive data. 
Typical studies in this field are mostly based on surveys, limited in scale and have self-reporting biases.

\section{Related Work}

\textbf{Celebrity on social media}:
A para-social relationship~\cite{horton1956mass} (PSR) describes imagined relationships that people have with a media persona. 
Before the advent of social media, PSR was mostly (i) one way, and (ii) through television. In recent years, the emergence of social media provided opportunities for increased interaction between celebrities and followers that did not previously exist. Such changes have the potential to make PSR more reciprocal and `social', or at least to intensify the illusion of sociability by creating seemingly more personal and frequent communication~\cite{pearson2010fandom}.
Celebrity interactions with fans by posting information about their daily routines and replying directly to fans have become increasingly common these days. Social media is being used as a way of self promotion for the celebrities, which change the way PSR are established and operate~\cite{marwick2011see,stever2013twitter}. 
Understanding celebrity behavior and fan reactions on social media has an impact on building good consumer relationships with brands~\cite{chung2014parasocial} 
and loyalty to the social networks~\cite{tsiotsou2015role}.

\textbf{Fandom on social media}:
Though most analysis of celebrities on social media to date have focused on how the celebrities place themselves, there are a few studies on the role fans' in interacting with celebs.
\cite{stavros2014understanding} study fans of NBA teams on Facebook and find four key motives for fans to engage with their favorite teams: `passion, hope, esteem and camaraderie'.
Kim et al.\cite{kim2016celebrity} study the role of celebrity self disclosure on Twitter and the reaction of fans. They report that increased celebrity self-disclosure leads to `enhanced fans' feeling of social presence, thereby positively affecting para-social interaction with celebrities'.

\textbf{Para-social breakups}: \cite{eyal2006good} introduce the idea of a para-social breakup, the end of a PSR. They find that a para-social breakup is like a regular romantic relationship breakup but not as intense.
Breakups on social media have been studied in the past~\cite{garimella2014love}. We define a para-social breakup on Twitter as an act of unfollowing the celebrity and study the impact of fan behavior on a breakup.
Work has been done on reasons for unfollowing others on Twitter~\cite{kwak2011fragile}. Kwak et al.\ conclude that most users unfollow `those who left many tweets within a short time, created tweets about uninteresting topics, or tweeted about the mundane details of their lives.'. 
In our case of a para-social breakup, we find that most unfollowing happens as 
an act of defiance against the celeb and despair of not receiving the personal attention that fans perceive they deserve.

\section{Dataset}
\label{sec:dataset}
We started with a set of the 15 most followed celebrities from popular culture on Twitter (in May 2015), from \url{http://followerwonk.com/bio}.
The selected celebrities were: 
@justinbieber, @katyperry, @taylorswift13, @ladygaga, @rihanna, @jtimberlake, 
@theellenshow, @kimkardashian, 
@cristiano,
@britneyspears, @jlo, @shakira, @selenagomez, @arianagrande, and @ddlovato. 
Note that we did not include politician @barackobama, as we assume political fandom to have very different characteristics.

Each of these celebrities has tens of millions of Twitter followers (fans). Since it is not feasible to analyze all their followers, we sampled a subset of fans for each celebrity. We sampled from three types of fans, defined based on their level of interaction with the celebrity. 

(1) Involved -- This is the set of fans who regularly interact with the celebrity in an intimate manner. To get this set, we first obtained all users replying to the tweets from celebrities and obtained users who replied to tweets from celebrities at least on five tweets with messages containing `i love you' (or similar variations, like, ILY).\footnote{This is common for many celebrities on Twitter. As an example, see any tweet from Justin Bieber \url{https://twitter.com/justinbieber/status/809785232186478592}} 

(2) Casual -- This is a set of fans who interacted with (replied /mentioned/retweeted) the celebrity at least once in the previous year (May 2014 -- May 2015). 

(3) Random -- Random sample of followers of the celebrity, sampled randomly from the millions of followers the celebrity has. 
These fans need not have interacted with the celebrity beyond the act of following.

The choice for these three sets was made in order to obtain the complete spectrum of fans, from the highly invested fans, who engage with the celebrity all the time, to people who just follow the celebrity out of curiosity.
We first randomly sampled approximately 2,000 users from each set for each celebrity. We then applied simple heuristics to remove bots and inactive accounts (at least 10 tweets/followers/friends, been on Twitter for at least one year, etc.). We then up and down sampled each group for each celeb to be approximately of the same size. This gave us 57,609 fans in total.

For these fans, we extracted a set of features based on a range of user interactions on Twitter. 
In an attempt to be as encompassing as possible, we did not apply strict theoretical considerations in identifying these features. Rather, we included any feature which may be at all relevant to fandom behavior and that we could quantify.  
We then allowed the factor analysis procedure (see next section) to group these as a way to reflect the underlying structure of the variables.

	(i) feat1 - Number of tweets,
        (ii) feat2 - Number of friends,
        (iii) feat3 - Number of followers,
        (iv) feat4 - Does the profile description contain an @mention of the celebrity?,
        (v) feat5 - Days since joining Twitter,
	(vi) feat6 - Fraction of tweets containing retweets,
        (vii) feat7 - Fraction of tweets containing mentions,
        (viii) feat8 - Fraction of tweets containing replies,
        (ix) feat9 - Fraction of tweets containing images,
        (x) feat10 - Fraction of tweets containing urls,
        (xi) feat11 - Fraction of tweets containing mentions of a celeb,
        (xii) feat12 - Fraction of tweets containing retweets of a celeb,
	(xiii) feat13 - Is the fan followed back by the celeb?,
        (xiv) feat14 - Fraction of followers followed by the celeb,
        (xv) feat15 - Fraction of friends followed by the celeb,
        (xvi) feat16 - Fraction of friends who also follow the same celeb,
        (xvii) feat17 - Fraction of followers who also follow the same celeb.

For all the users, we used Face++ Api\footnote{\url{http://www.faceplusplus.com/}} on their profile pictures to obtain estimates of their age and gender. 
Face++ uses computer vision and data mining techniques applied to a large database of celebrities to generate estimates of age and sex of individuals from their pictures.
We were able to obtain confident age and gender predictions for around 50\% of the users (27,889 users).

\section{Identifying Factors of Fandom}
Since having 17 features makes it hard to interpret the underlying dynamics of the features clearly, we applied factor analysis to find a smaller number of factors that capture the correlations between the features.
Factor analysis is a simple statistical method~\cite{harman1960modern} to first find correlations among variables in order to then find a lower number of unobserved variables called factors.

All the feature variables were first z-normalized (to have zero mean and standard deviation one).
We create a matrix where the rows are the fans and columns are the 17 features (of size\ 57,609 $\times$ 17).
We then computed pairwise correlations between the 17 features, 
and ran a factor analysis on this matrix.
Using a cut off for the eigenvalue of 1.25, we obtained three factors that explain about 57\% of the variance (as proposed in \cite{yong2013beginner}). The three main factors we inferred are:

\begin{squishlist}
	\item Factor1: \{feat4, feat11, feat12\}. Looking at the features that load well with this factor, we define this factor to indicate the \textit{Devotedness} of a fan.
	\item Factor2: \{feat13, feat14, feat15\}. These set of features seem to indicate the \textit{Sociability} of fandom. 
	\item Factor3: \{feat1, feat2, feat3, feat5\}, which indicates the \textit{Activity} of the fan on Twitter.
\end{squishlist}

Note that we were able to identify these feature groups (factors) in a completely unsupervised manner.

\textbf{Quantifying fandom}: Using the three factors obtained above, we are now able to get a concise representation of the features that encode fandom on Twitter.
For each fan, we compute three scores, 
\begin{gather}
	Devotedness = median(feat4, feat11, feat12) \\
	Sociability = median(feat13, feat14, feat15) \\
	Activity = median(feat1, feat2, feat3, feat5)
\end{gather}
Note that since the features are z-normalised, the absolute values of these scores do not give much information, but relative trends, comparing different scores do.

\begin{figure}
\centering
\includegraphics[width=\columnwidth]{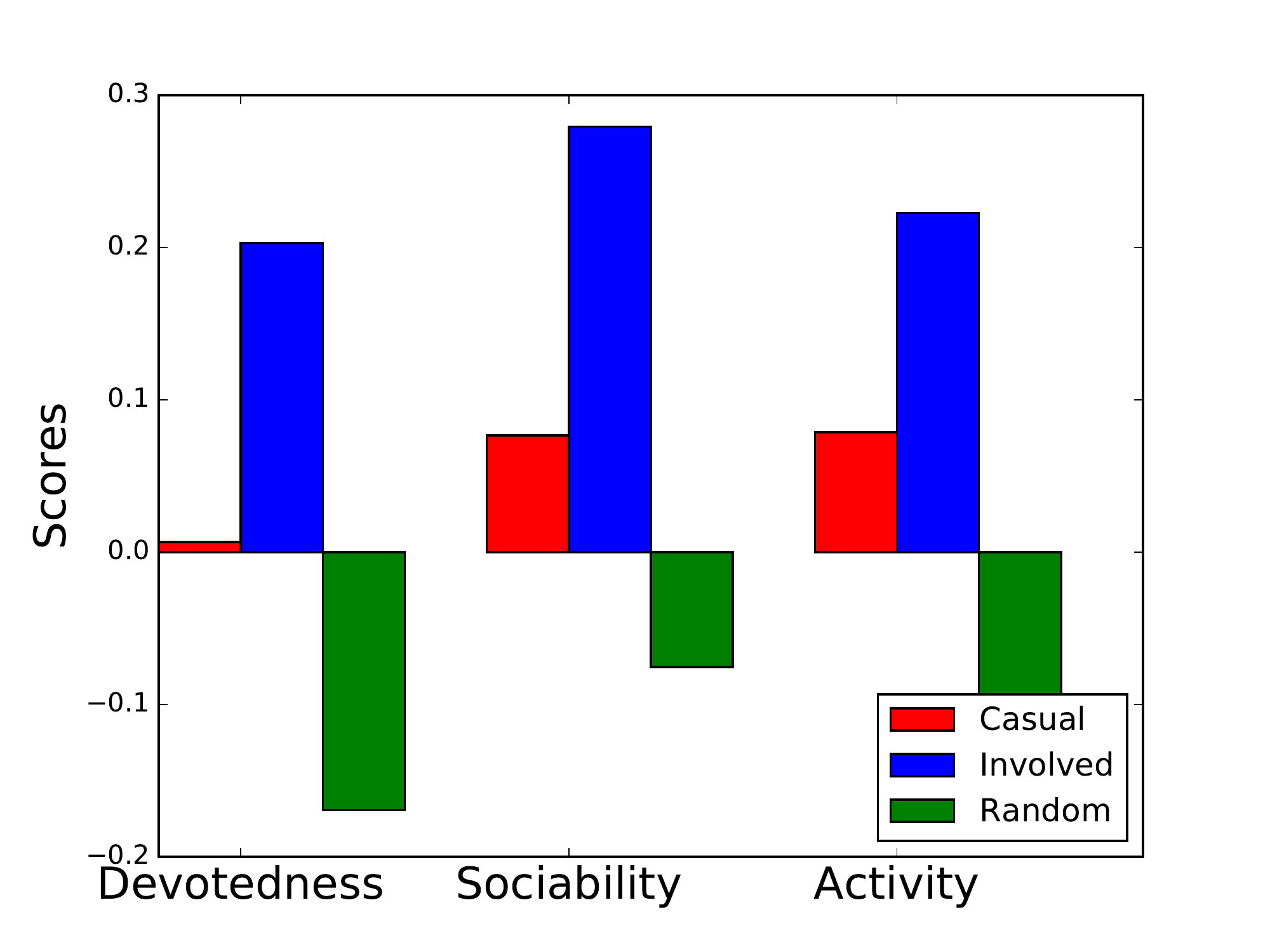}
\caption{Factor values across fan types.}
\label{fig:barplot}
\vspace{-\baselineskip}
\end{figure}

\textbf{Validation}: 
To validate our factor analysis we tested whether the three factors could reconstruct the fan types of our sample that were discerned based on other criteria, such as frequency of tweets, intimacy of interactions with the celebrity, etc.
The results are shown in Figure~\ref{fig:barplot}. The figure indicates that involved fans have a high devotedness, activity and sociability compared to the other groups, which is in line with the way we defined the groups. 
This is non obvious as the type of attributes we  consider to obtain the factors was defined in a generic way, without taking into account the fan type.

Next, we computed correlations between the various factors for all the fans. We find that devotedness and sociability are highly correlated (Pearson's R 0.65, p$<$1e-6). We do not find meaningful correlations between devotedness/sociability and activity.

\textbf{Differences in gender and age group}:
Using data from the 27,000 fans for whom we could get gender and age data, we tried to see if there are any visible differences between genders in terms of the three factors.
We compared the mean values of the factors for the two groups (men and women) and there are significant differences. Using a Mann-Whitney U test, we find that women have a higher devotedness and sociability, while men have higher activity (p$<$0.0001).

Next, we bucketed the age estimates into 4 groups - 0--14, 15--25, 26--35 and 36+.
We computed the mean devotedness, sociability and activity values for all fans in each age group.
We find that younger fans (from the 0--14 and 15--25 buckets) are more devoted, social and active (statistical significance tested using a Kruskals-Wallis test, 
p$<$1e-6).
Detailed results are omitted due to lack of space.

\section{Para-social Breakups}
In this section, we try to use the fan behavior characteristics we obtained to understand para-social breakups. 
We assume that a para-social breakup on Twitter manifests itself as an act of unfollowing the celebrity.
We tracked all the 57k fans for a period of 26 weeks (between 21 May 2015 -- 21 Nov 2015), and got data on whether they still follow the celebrity every week. At the end of the data collection period, we recorded 2,369 fans unfollowing a celebrity during this period. 
We also estimate when the fans started following the celebrity using \cite{meeder2011we}.

As a first step, we tried to predict if a user would unfollow or not, just based on the three factors as features. Since unfollowing is a rare event, we first over sample from the rare group using SMOTE \cite{Chawla:2002:SSM:1622407.1622416}. Then, using a random forest classifier, and 10-fold cross-validation, we were able to predict if a user will unfollow or not with a 66\% accuracy. This may not look like a great accuracy, but it is still much better than random. Given that we are only using three features (the three factors), the factors seem to capture some signal in the unfollowing behavior.

Next, we model unfollowing behavior using Survival analysis.
Survival analysis is a statistical tool for analyzing the expected time to an event (unfollowing, in this case). It can be used to answer questions such as: whether a group of fans is more likely to unfollow than others.
We set up survival analysis as follows: (i) \emph{Event}: The act of unfollowing a celeb, (ii) \emph{Survival time/event time}: Months since following the celebrity to unfollow. e.g. if a fan follows a celeb in Jan 2015 and unfollows in Oct 2015, this variable is 10. We have data going back to the last 4 years (48 months).
(iii) \emph{Censoring event}: All users who haven't unfollowed yet. For example, if a user followed the celebrity in Jan 2015 and is still following, we censor the user after 23 months (Jan 2015 to Dec 2016)
(iv) \emph{Survival function}: Probability of unfollowing after $x$ months. 

\begin{figure}
\centering
\includegraphics[width=0.8\columnwidth]{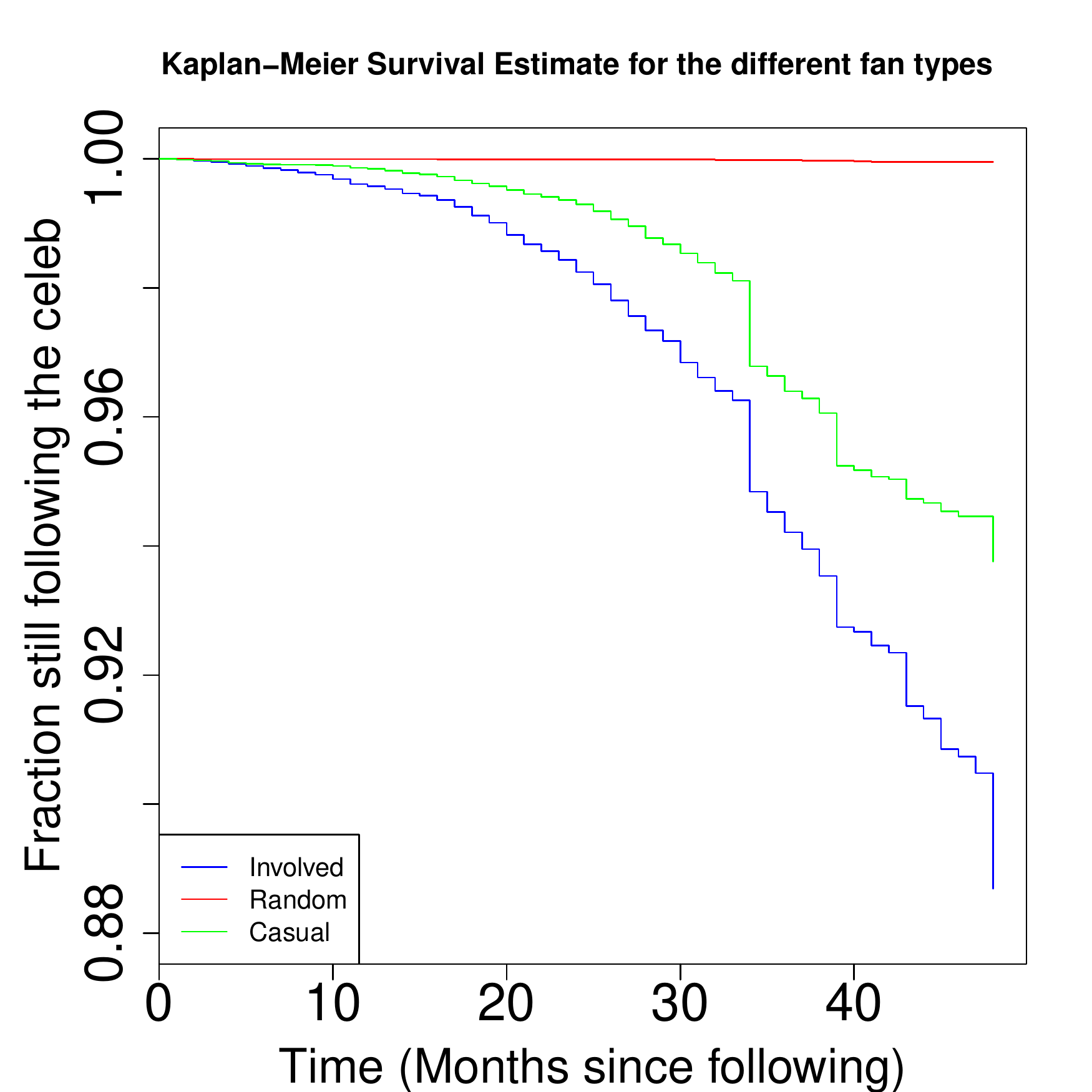}
\caption{Survival probability for the three fan types\label{fig:survival1}}
\vspace{-\baselineskip}
\end{figure}

Given this set up, our first task is to see if there is any difference in para-social breakups between the three fan types (involved, casual and random).
Figure~\ref{fig:survival1} shows the survival probability for the various fan types using a Kaplan Meier non-parametric analysis. 
Compared across groups, involved users have a statistically significantly higher  probability of unfollowing. We can also see that the chance of unfollowing for random users is almost non-existent.

This finding might be counter intuitive, given we expect involved fans to feel ``closer'' to the celebrity. We can interpret this behavior in terms of a cost/benefit analysis. The more a fan engages the celeb, the more likely it is for the relationship to end. The more involved a fan is the more the emotional cost they invest in the relationship and thus the more likely they are to end it when it does not reward them. 
These high investors tend to be young and female whereas as older fans are less invested and less prone to disappointment and breakup. 

We manually looked at the profile descriptions of users who unfollowed, before and after unfollowing, and found interesting examples. 
(i) Before: `@arianagrande followed and faved (30.11.14)', After: `dreams dont work unless you do. good vibesss!!'
(ii) Before: `thank you justin my life', After:  `find your purpose'
(iii) Before: `queen faved x2. pls follow 2, ma queen', After: `im loving the pain, i never wana live without it'.
These examples show cases of despair and a sense of defiance after not getting the personal attention they desired.

Next, we compared the unfollowing behavior with respect to the factor scores obtained for each fan. Since these values are continuous (not categorical), we applied a Cox proportional hazard regression analysis to obtain the coefficients for the three factors. 
The coefficients for sociability and activity were statistically significant (p $<$ 0.001 ), with values, 0.421 and 0.418 respectively.
Using the exponent of the coefficient ($e^{coeff}$), we can conclude that users that have higher sociability and activity, will have a lower unfollowing time, i.e.\ they will unfollow earlier. 

This could mean that when fandom is not wholly directed at the celebrity but rather also has a communal component, it is more stable. Going back to the cost/benefit perspective we offered above, we can say that the socialbility of fandom adds to the benefits fans receive from their PSRs so that even if a celebrity disappoints on Twitter or in real life, the community of fans provides benefits to its members such that they are less likely to break off this relationship.

\section{Conclusions}
In this paper, we presented a large scale analysis of fan behavior on Twitter. 
Based on fan-celeb interaction and fan activity, we first obtained the important factors that underlie fan behavior. 
Using these factors, we analysed (i) fan behavior by gender \& age, and (ii) para-social breakup behavior. 
Our results show that (i) fandom is a social phenomenon, (ii) female fans are often more devoted and younger fans are more active \& social, and (iii) the most devoted fans are more likely to be involved in a para-social breakup.

In summary, our analyses point to the complexity that social media use adds to fandom behavior and the various ways that fans develop PSRs on Twitter. As celebrities - and now presidents -continue to use social media as a central avenue to cultivate their following, engage their fans and influence society, it is crucial that more studies are conducted to understand the ways such relationships develop and are maintained.  

\section{Acknowledgements.}
This work has been partly supported by the Academy of Finland project ``Nestor'' (286211) and the EC H2020 RIA project ``SoBigData'' (654024).

\bibliographystyle{ACM-Reference-Format}
\bibliography{related}

\end{document}